\documentclass[aps,showpacs]{revtex4-1}
\usepackage{amssymb}
\usepackage{graphicx}

\begin{document}

\title{Three-dimensional hybrid vortex solitons}
\date{\today}
\author{Rodislav Driben$^{1}$}
\author{Yaroslav V. Kartashov$^{2,3}$}
\author{Boris A. Malomed$^{4}$}
\author{Torsten Meier$^1$}
\author{Lluis Torner$^{2}$}

\affiliation{$^1$Department of Physics \& CeOPP, University of
Paderborn, Warburger Str. 100,
Paderborn D-33098, Germany \\
$^2$ICFO-Institut de Ciencies Fotoniques, and Universitat
Politecnica de Catalunya,
Mediterranean Technology Park, E-08860 Castelldefels (Barcelona), Spain\\
$^3$Institute of Spectroscopy, Russian Academy of Sciences, Troitsk,
Moscow, 142190, Russia\\
$^4$Department of Physical Electronics, School of Electrical
Engineering, Faculty of Engineering, Tel Aviv University, Tel Aviv
69978, Israel}

\begin{abstract}
We show, by means of numerical and analytical methods, that media with a
repulsive nonlinearity which grows from the center to the periphery support
a remarkable variety of previously unknown complex stationary and dynamical
three-dimensional solitary-wave states. Peanut-shaped modulation profiles
give rise to vertically symmetric and antisymmetric vortex states, and novel
stationary \textit{hybrid states}, built of top and bottom vortices with
opposite topological charges, as well as robust \textit{dynamical hybrids},
which feature stable precession of a vortex on top of a zero-vorticity base.
The analysis reveals stability regions for symmetric, antisymmetric, and
hybrid states. In addition, bead-shaped modulation profiles give rise to the
first example of exact analytical solutions for stable three-dimensional
vortex solitons. The predicted states may be realized in media with a
controllable cubic nonlinearity, such as Bose-Einstein condensates.
\end{abstract}

 \maketitle

\section{Introduction}

Self-trapping of three-dimensional (3D) confined modes (solitons or, more
properly, solitary waves) in optics \cite{review1,Dum,review2},
Bose-Einstein condensates (BECs) \cite{BEC-skyrm,BEC-vort,Delgado},
ferromagnetic media \cite{ferro-vortring}, superconductors \cite%
{superconductor-knot}, semiconductors \cite{semi}, baryonic matter \cite%
{low-energy}, and general field theory \cite{fields,fields-review} is a
fundamental problem of nonlinear physics. An apparent condition is that an
attractive, or self-focusing, nonlinearity is necessary for the creation of
localized states; however, the attractive cubic nonlinearity simultaneously
gives rise to collapse \cite{Kuznetsov} of localized modes in
higher-dimensional settings and, additionally, to strong azimuthal
modulational instability of states with intrinsic vorticity \cite{Anton},
thus making the search for stable 3D fundamental and topological solitons in
materials with the cubic (Kerr) nonlinearity a challenging issue.

Various methods have been elaborated over the years, chiefly in the
theoretical form, to remedy this situation and stabilize 3D solitary waves,
fundamental and vortical ones alike. As outlined in detail in the reviews
\cite{review1,Dum} (see also the more recent work \cite{CQ}), stabilization
may be achieved by a higher-order quintic self-defocusing nonlinearity,
provided that the underlying physical setting gives rise to such terms.
Another possibility is offered by periodic (lattice) potentials \cite%
{review1,Dum,review2}. In particular, a 2D potential may be
sufficient for the stabilization of 3D solitons, as well as for the
stabilization against supercritical collapse \cite{quasi-2D}. In
addition, it is also possible to stabilize 3D fundamental solitons
by means of ``nonlinearity management" (time-periodic sign-changing
modulation of the nonlinearity coefficient), which should be
combined, at least, with a 1D lattice potential \cite{Marek}. The
use of nonlocal nonlinearities may also help to stabilize 3D
localized modes \cite{nonlocal}. Lastly, it is relevant to mention a
very recent result concerning 2D localized modes created by the
self-focusing cubic nonlinearity in the free space: while a common
belief was that they might never be stable, it has been demonstrated
in Ref. \cite{HS} that mixed vortex-fundamental modes in a system of
two coupled GP equations modeling the spin-orbit-coupled BEC can be
{\em stable} in the 2D free space. This unexpected result is
explained by the fact that the norm of the mixed modes takes values
below the well-known 2D-collapse threshold \cite{Kuznetsov}.

Unlike the above-mentioned methods, the use of spatially inhomogeneous cubic
nonlinearity does not yield stabilization of 3D solitons \cite{review2}. In
the 2D setting, a nonlinearity subject to a smooth spatial modulation cannot
stabilize solitons either \cite{Sivan}. Stabilization of 2D fundamental
solitons (but not vortex states) is possible by means of various spatial
modulation profiles with sharp edges \cite{2D-stabilization}. For this
reason, most of previous studies of solitons in inhomogeneous nonlinearity
landscapes have been performed in 1D settings, chiefly for periodic
modulation patterns \cite{aa}.

A radically different approach was recently put forward and
elaborated in Refs. \cite{we} and \cite{gyro}: a \emph{repulsive},
or defocusing, nonlinearity, whose local strength grows from the
center to the periphery, as a function of radius $r$ at any rate
faster than $r^{3}$, can readily induce self-trapping of robust
localized modes, which are stable not only to weak, but also to
strong perturbations (although these solutions are far from those in
integrable models, we call them ``solitons", as commonly adopted in
the current literature when dealing with stable self-trapped modes).
In BECs, the necessary spatial modulation of the nonlinearity may be
induced by means of the tunable Feshbach resonance,
controlled by magnetic \cite{extreme-tunability} and/or optical \cite%
{optical-Feshbach} fields, created with appropriate inhomogeneous profiles
\cite{Feshbach}. The required magnetic field patterns can be provided by
magnetic lattices of various types \cite{magnetic-lattice}, while the
optical-intensity profiles can be \textit{painted} by laser beams in 3D
geometries \cite{paint}. In addition to fundamental solitons, landscapes
with a growing repulsive nonlinearity were shown to support topological
states in the form of vortex-soliton tori, which can exhibit gyroscopic
precession under the action of an external torque \cite{gyro} (precession of
a tilted vortex was earlier considered in a different setting in Ref. \cite%
{precession}).

So far, only the simplest 3D vortex solitons were addressed in the framework
of the setting based on the spatially modulated strength of the
self-repulsion. The possibility of the existence of more complex
vorticity-carrying 3D structures remains unexplored. In this context, it
should be stressed that the creation of stable structures carrying several
topological dislocations is a complex challenge. Previously, such entities
were found mostly in the form of vortex-antivortex pairs and vortex arrays
in settings with a reduced dimensionality, such as superconductors \cite%
{vort-dipole-supercond,multivort-dipole-supercond}, pancake-shaped atomic
Bose-Einstein condensates \cite{vort-dipole-BEC,multivort-dipole-BEC}, and
exciton-polariton condensates \cite{vort-dipole-exciton-polariton}. To the
best of our knowledge, no examples of 3D solitons with coaxial vortex lines
threading several objects forming a complex state, or with the topological
charge changing along the axis of the soliton, have been reported.

In this work, our analysis reveals that 3D media with a repulsive
nonlinearity growing from two symmetric minima to the periphery make it
possible to create complex but, nevertheless, stable static and dynamical
self-trapped topological modes, in the form of fundamental and vortical
dipoles, stationary vortex-antivortex hybrids, and precessing hybrids built
as a vortex sitting on top of a zero-vorticity mode. These are remarkable,
novel species of 3D localized modes, which have not been reported before in
any other systems. The very existence of the stationary vortex-antivortex
solitons and precessing vortex-fundamental hybrids is an unexpected finding,
because the topology of such states is different in their top and bottom
sections. All these previously unknown static and dynamical states are
supported by the nonlinearity-modulation profile, which is obtained from the
spherical configuration by a deformation in the axial (vertical) direction.

The basic model is introduced in Section II, where we also give a number of
analytical results, which can be obtained in spite of the apparent
complexity of the system. These include the Thomas-Fermi approximation (TFA)
for families of vortex modes, an approximate description of the dipole
(antisymmetric) modes in terms of quasi-1D dark solitons embedded into the
ordinary symmetric states, and an approximation which explains the existence
of stationary vortex-antivortex hybrids. Results of systematic numerical
analysis are reported in Section III, including families of stationary
antisymmetric and vortex-antivortex hybrid modes, as well as dynamical
(steadily precessing) vortex-fundamental hybrids. A comprehensive stability
analysis is presented too, along with simulations of the spontaneous
evolution of unstable states. The work is concluded by Section IV. In the
Appendix, we additionally present stable analytical solutions for 3D vortex
solitons in a model with a bead-shaped spatial modulation profile, which is
the first example of any system admitting exact solutions of this type, thus
providing a direct proof of their existence.

\section{The models and analytical results}

\subsection{The general formulation}

Our system is described by the single-component nonlinear Schr\"{o}%
dinger/Gross-Pitaevskii (NLS/GP) equation in the 3D space for the wave
function $\psi (\mathbf{r},t)$:
\begin{equation}
i\frac{\partial \psi }{\partial t}=-\nabla ^{2}\psi +\sigma (\mathbf{r}%
)\left\vert \psi \right\vert ^{2}\psi ,  \label{1}
\end{equation}%
where Laplacian $\nabla ^{2}$ acts on coordinates $\mathbf{r}=\left\{
x,y,z\right\} $, and $\sigma (\mathbf{r})>0\ $represents the local strength
of the repulsive nonlinearity, which must grow at $r\rightarrow \infty $
faster than $r^{3}$. Dynamical invariants of Eq. (\ref{1}) are the norm and
Hamiltonian, $N=\int \int \int \left\vert \psi \left( x,y,z,t\right)
\right\vert ^{2}dxdydz$ and $H=\int \int \int \left[ \left\vert \nabla \psi
\right\vert ^{2}+(1/2)\sigma (r)|\psi |^{4}\right] dxdydz$. Stationary
states with real chemical potential $\mu $ can be found in the form of $\psi
\left( \mathbf{r},t\right) =\exp \left( -i\mu t\right) \phi (\mathbf{r})$,
where the (generally, complex) spatial wave function satisfies the equation
\begin{equation}
\mu \phi =-\nabla ^{2}\phi +\sigma (\mathbf{r})\left\vert \phi \right\vert
^{2}\phi .  \label{2}
\end{equation}

While the simplest 3D vortex solitons have been obtained in
spherically-symmetric nonlinearity landscapes, such as the one with $\sigma
(r)=\exp \left( r^{2}/2\right) $ \cite{gyro}, here our objective is to a
show that a deformation of this nonlinearity profile, lending it two local
minima, allows us to produce novel species of robust stationary and
precessing 3D topological modes. To this end, the spherically symmetric
modulation pattern is shifted by the distance $\pm d/2$ along the $z$ axis,
and the so produced profiles are stitched together in the midplane, $z=0$:
\begin{equation}
\sigma \left( \rho ,z\right) =\exp \left[ \frac{1}{2}\left( \rho ^{2}+\left(
|z|-\frac{d}{2}\right) ^{2}\right) \right] ,  \label{sigma}
\end{equation}%
with $\rho ^{2}\equiv x^{2}+y^{2}$. This profile keeps the cylindrical
symmetry and, accordingly, the $z$-component of the field's angular
momentum, which is the third dynamical invariant of the model, in addition
to $N$ and $H$,
\begin{equation}
M=i\int \int \int \psi ^{\ast }\left( y\partial _{x}-x\partial _{y}\right)
\psi dxdydz,  \label{M}
\end{equation}%
where $\ast $ stands for the complex conjugation.

The steep anti-Gaussian profile, adopted in Eq. (\ref{sigma}), is not a
necessary feature of the model. As mentioned above, the necessary condition
for the existence of 3D solitons, which follows from the normalizability of
the wave function, is that $\sigma (r)$ must grow faster than $r^{3}$ \cite%
{we}. The modulation profile (\ref{sigma}) is adopted here as it makes it
possible to obtain families of stationary vortex modes in an almost exact
analytical form, by means of the TFA, thus supporting numerical findings.

\subsection{Symmetric self-trapped vortices and the Thomas-Fermi
approximation (TFA)}

Among the complex stable modes reported below, the simplest species are
confined vortex states, carrying an integer topological charge $S$, which
are looked for, in the cylindrical coordinates, as
\begin{equation}
\phi (\rho ,z,\theta )=\exp \left( iS\theta \right) \Phi \left( \rho
,z\right) ,  \label{S}
\end{equation}%
where $\Phi $ is a real function. As follows from Eqs. (\ref{M}) and (\ref{S}%
), the angular momentum of the vortex is $M=SN$. Below, such modes, with
identical vorticities $S$ in the top and bottom parts of the peanut-shaped
nonlinearity landscape, are denoted as $S/S$ (the definitions of ``top" and
``bottom" are arbitrary here, as Eqs. (\ref{2}) and (\ref{sigma}) are
obviously invariant with respect to $z\rightarrow -z$).

\begin{figure}[t]
\centering\centering\includegraphics[width=8.5cm]{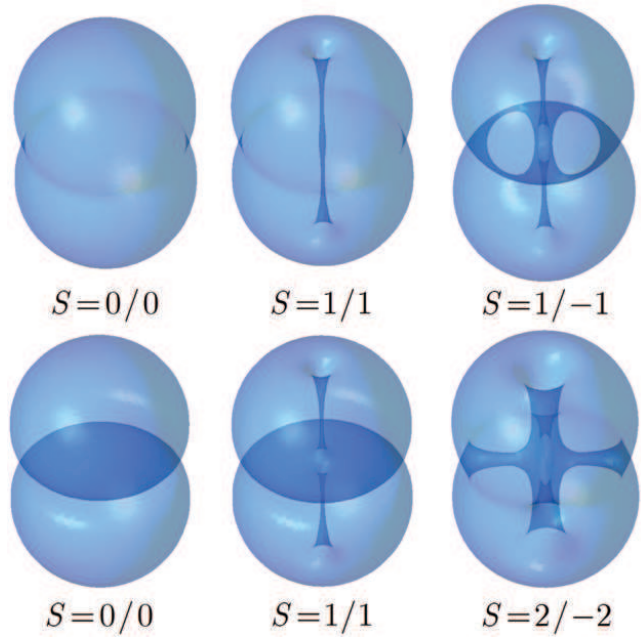}
\caption{(Color online) Three-dimensional images of modes are displayed by
means of isosurfaces corresponding to $\left\vert \protect\psi \left(
x,y,z\right) \right\vert ^{2}=0.2$. The vorticity content of the states is
indicated under each panel. For $S=0/0$ and $S=1/1$, the top and bottom
panels display the symmetric and antisymmetric varieties, respectively. The
states shown in the top row are stable, while those in the bottom row are
unstable. All the modes pertain to $d=3$ in Eq. (\protect\ref{sigma}) and $%
\protect\mu =10$, except for the one with $S=1/-1$, which was obtained for $%
d=5$ and $\protect\mu =7$.}
\label{fig1}
\end{figure}

\begin{figure}[t]
\centering\centering\includegraphics[width=9cm]{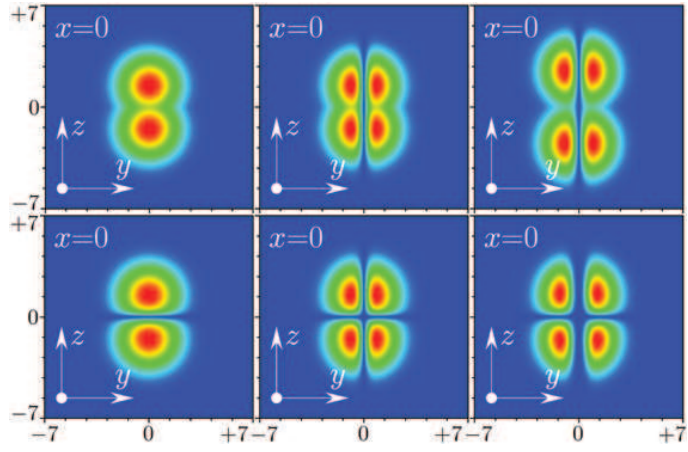}
\caption{(Color online) Density plots in vertical cross sections, $x=0$, of
the 3D modes displayed in the corresponding panels of Fig. \protect\ref{fig1}%
.}
\label{fig2}
\end{figure}

\begin{figure}[t]
\centering\centering\includegraphics[width=8cm]{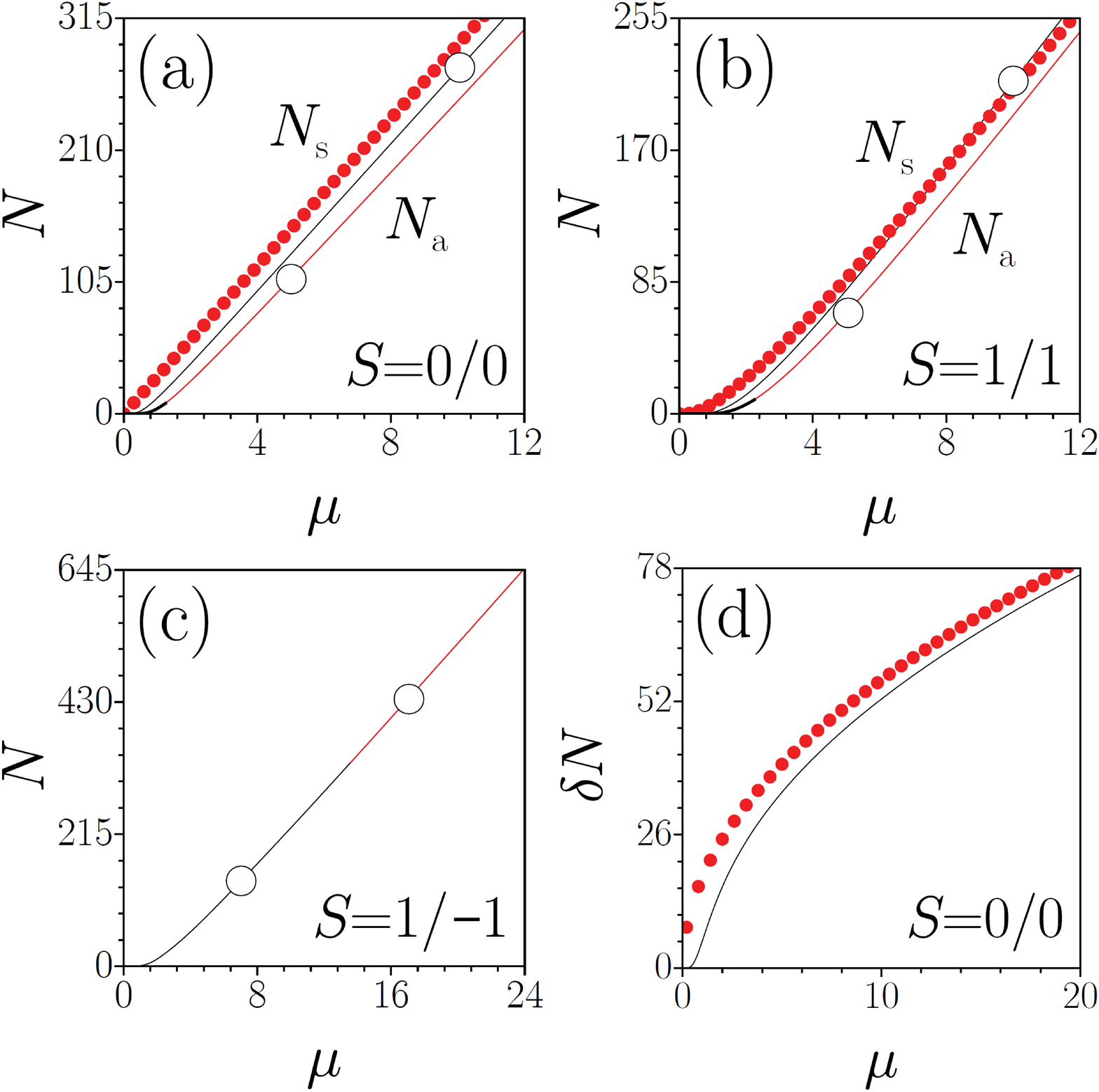}
\caption{(Color online) (a,b) Norms $N_{s}$ and $N_{a}$ of the symmetric and
antisymmetric varieties of the fundamental ($S=0/0$) and vortex ($S=1/1$)
modes, versus chemical potential $\protect\mu $, with $d=3$ in Eq. (\protect
\ref{sigma}). In these panels [as well as in (c)], black and red segments
designate stable and unstable (sub)families, respectively [the short black
(stable) segments in panels (a) and (b) are made bolder for better
visibility]. Chains of red dots represent the prediction of the Thomas-Fermi
approximation, as given by Eq. (\protect\ref{NTFA}. (c) The $N(\protect\mu )$
dependence for the \textit{hybrid mode} of the $1/-1$ type, with $d=5$ in
Eq. (\protect\ref{sigma}). Circles on stable (black) branches mark typical
examples of the stationary modes, which are displayed in the top rows of
Figs. \protect\ref{fig1} and \protect\ref{fig2}. The development of the
instability of the modes labeled by the circles on unstable (red) segments
is shown below in Fig. \protect\ref{fig6}. (d) The comparison of the
numerically found norm difference between the symmetric and antisymmetric
varieties of the fundamental mode of the $0/0$ type (the continuous line),
and the respective analytical approximation given by Eq. (\protect\ref{I})
(red dots), for $d=0$ in Eq. (\protect\ref{sigma}).}
\label{fig3}
\end{figure}

The shape of the simplest symmetric vortices and fundamental solitons ($S=0$%
) can be approximated by means of the TFA, which neglects $z$- and $\rho $-
derivatives in Eq. (\ref{2}), and is usually relevant in the case of a
strong repulsive nonlinearity \cite{TF-Pu,Poland,BEC-vort}:
\begin{widetext}
\begin{equation}
\Phi _{\mathrm{TFA}}^{2}(\rho ,z)=\left\{
\begin{array}{c}
0,~\mathrm{at}~~\rho ^{2}<\rho _{S}^{2}\equiv S^{2}/\mu , \\
\left( \mu -S^{2}/\rho ^{2}\right) \exp \left[ -\frac{1}{2}\left( \rho
^{2}+\left( |z|-\frac{d}{2}\right) ^{2}\right) \right] ,~\mathrm{at}~~\rho
^{2}>\rho _{S}^{2}.%
\end{array}%
\right.  \label{uTFA}
\end{equation}
\end{widetext}Here the first line represents the hole at the center of the
vortex state (see panels marked $1/1$ in the top rows of Figs. \ref{fig1}
and \ref{fig2}). Families of self-trapped modes are characterized by
dependence $N(\mu )$, which can be obtained from Eq. (\ref{uTFA}) in an
approximate analytical form:
\begin{widetext}
\begin{equation}
N_{\mathrm{TFA}}^{(S)}=4\pi \mu ^{2}e^{-S^{2}/\left( 2\mu \right)
}\int_{0}^{\infty }dR\frac{Re^{-R}}{\mu R+S^{2}/2}\int_{0}^{\infty
}dze^{-\left( z-d/2\right) ^{2}/2}.  \label{NTFA}
\end{equation}
\end{widetext}For $S=0$ (the fundamental mode), Eq. (\ref{NTFA}) reduces to
a simple linear dependence, $N_{\mathrm{TFA}}^{(S=0)}=2\sqrt{2\pi ^{3}}\left[
1+\mathrm{erf}\left( d/2\sqrt{2}\right) \right] \mu .$ The constant slope $%
dN/d\mu $ given by the latter expression is, actually, an asymptotically
exact result at large $\mu $ for any $S$. Figures \ref{fig3}(a,b) show that,
while the TFA predictions for $N(\mu )$ may be shifted from their
numerically found counterparts, the asymptotic slope is indeed predicted
exactly.

Our stability analysis for vortex modes (\ref{S}), as well as for other
stationary modes featuring the cylindrical symmetry, which are considered
below, was carried out by numerically solving the linearized equations for
small perturbations. Perturbed solutions are sought for as
\begin{widetext}
\begin{equation}
\psi \left( \rho ,z,t\right) =e^{-i\mu t+iS\theta }\left\{ \Phi \left( \rho
,z\right) +\epsilon \left[ e^{ik\theta +\delta t}\varphi _{+}\left( \rho
,z\right) +e^{-ik\theta +\delta ^{\ast }t}\varphi _{-}^{\ast }\left( \rho
,z\right) \right] \right\} ,  \label{pert}
\end{equation}
\end{widetext}where $\epsilon $ is an infinitesimal amplitude of the
perturbation, $k$ is its integer azimuthal index, and $\delta \left( S,\mu
,k\right) $ is a (generally, complex) instability growth rate. Substitution
of expression (\ref{pert}) into Eq. (\ref{1}) and the linearization gives
rise to the eigenvalue problem for $\delta $ represented by the following
equations:
\begin{widetext}
\begin{eqnarray}
\left( \mu +i\delta +\frac{\partial ^{2}}{\partial \rho
^{2}}+\frac{1}{\rho } \frac{\partial }{\partial \rho
}-\frac{\left( S+k\right) ^{2}}{\rho ^{2}}+ \frac{\partial
^{2}}{\partial z^{2}}\right) \varphi _{+} &=&\sigma \left( \rho
,z\right) \Phi ^{2}\left( \rho ,z\right) \left( 2\varphi
_{+}+\varphi
_{-}\right) ,  \nonumber \\
\left( \mu -i\delta +\frac{\partial ^{2}}{\partial \rho
^{2}}+\frac{1}{\rho } \frac{\partial }{\partial \rho
}-\frac{\left( S-k\right) ^{2}}{\rho ^{2}}+ \frac{\partial
^{2}}{\partial z^{2}}\right) \varphi _{-} &=&\sigma \left( \rho
,z\right) \Phi ^{2}\left( \rho ,z\right) \left( 2\varphi
_{-}+\varphi _{+}\right) .  \label{eigen}
\end{eqnarray}
\end{widetext}The stability condition is $\mathrm{Re}\left\{ \delta \left(
S,\mu ,k\right) \right\} =0$, which must hold for all eigenvalues at given
values of $S$ and $\mu $.

\subsection{Dipole (antisymmetric) modes}

The vortex and fundamental modes can be \textit{twisted} in the vertical
direction, which lends them an antisymmetric (dipole) structure along the $z$
axis, as depicted in the left and middle panels in the bottom rows of Figs. %
\ref{fig1} and \ref{fig2}. Dipole modes have been previously studied in
diverse 1D and 2D settings \cite{dipole-soliton}, including vortex dipoles
created in a common plane \cite%
{vort-dipole-supercond,vort-dipole-BEC,vort-dipole-exciton-polariton}. In
3D, such dipole structures can be approximately described by assuming that a
quasi-1D dark soliton is embedded into an originally symmetric 3D mode
around its midplane ($z=0$), as suggested in a different context in Ref.
\cite{Delgado}. In particular, for the fundamental states ($S=0$)
approximated by the TFA expression (\ref{uTFA}), the respective
antisymmetric solution can be easily found from Eq. (\ref{2}), assuming that
the width of the dark soliton in the $z$ direction is much smaller than the
intrinsic scale of the TFA mode, i.e., $\mu $ is large enough:
\begin{widetext}
\begin{equation}
\Phi _{\mathrm{anti}}\left( \rho ,z\right) =\sqrt{\mu }\exp \left[
-\frac{1}{ 4}\left( \left( \frac{d}{2}-|z|\right) ^{2}+\rho
^{2}\right) \right] \tanh \left[ \sqrt{\mu /2}e^{-\left(
d/4\right) ^{2}}z\right] .  \label{DS}
\end{equation}
\end{widetext}For the vortex states, a similar approximation is available
too, but its applicability condition does not hold around the inner hole of
the vortex.

Solution (\ref{DS}) corresponds to a gap which cleaves the antisymmetric
mode around $z=0$, as shown in the bottom row of Figs. \ref{fig1} and \ref%
{fig2}. The width of the gap does not depend on $\rho $, implying that the
gap is \emph{nearly flat}, which is well corroborated by numerical results,
see the left panel in the bottom row of Fig. \ref{fig2}. Solution (\ref{DS})
makes it possible to calculate the difference between the norm of the
symmetric state and its antisymmetric counterpart. Indeed, Eqs. (\ref{uTFA})
and (\ref{DS}) yield
\begin{widetext}
\begin{equation}
\delta N(\mu )=2\pi \int_{-\infty }^{+\infty }dz\int_{0}^{\infty
}\rho d\rho \left[ \Phi _{\mathrm{TFA}}^{2}\left( \rho
,x;S=0\right) -\Phi _{\mathrm{anti }}^{2}\left( \rho ,x\right)
\right] =4\pi \sqrt{2\mu }e^{-\left( d/4\right) ^{2}}.  \label{I}
\end{equation}
\end{widetext}As shown in Fig. \ref{fig3}(d), this prediction is quite
accurate.

\subsection{Hybrid modes}

Completely novel species of stationary 3D modes are \textit{hybrids} of the $%
S/-S$ type, which combine vortex states with opposite signs and equal norms
in the top and bottom sections of the peanut-shaped structure, as shown in
Figs. \ref{fig1} and \ref{fig2}. Unlike the symmetric and antisymmetric
vortices introduced above, the hybrids cannot feature axisymmetric density
distributions. A central question is whether the vortex-antivortex hybrids
exist as stationary modes and, if they do, whether they can be stable. To
address this issue, a stationary solution may be looked for in an
approximate form as%
\begin{equation}
\phi \left( \rho ,\theta ,z\right) =\phi _{+}\left( \rho ,z\right)
e^{iS\theta }+\phi _{-}\left( \rho ,z\right) e^{-iS\theta },  \label{hybrid}
\end{equation}%
assuming that $\phi _{+}\left( \rho ,z\right) $ and $\phi _{-}\left( \rho
,z\right) $ rapidly vanish, respectively, at $z<0$ and $z>0$, so that the
two vortical components form a sharp \textit{domain wall} close to $z=0$.
Substituting ansatz (\ref{hybrid}) in Eq. (\ref{2}), and using the
rotating-wave approximation, one arrives at a system of nonlinearly coupled
equations,
\begin{widetext}
\begin{equation}
\left( \mu +\frac{\partial ^{2}}{\partial \rho ^{2}}+\frac{1}{\rho
}\frac{
\partial }{\partial \rho }-\frac{S^{2}}{\rho ^{2}}+\frac{\partial ^{2}}{
\partial z^{2}}\right) \phi _{\pm }=\sigma \left( \rho ,z\right) \left(
2\phi _{\mp }^{2}+\phi _{\pm }^{2}\right) \phi _{\pm }~.  \label{wall}
\end{equation}
\end{widetext}Note that in the right-hand sides of this equation the
cross-phase-modulation coefficient is twice as large as its
self-phase-modulation counterpart. This is typical for systems which give
rise to solutions in the form of sharp domain walls between states with
different wave numbers, linear or azimuthal ones \cite{GL,Poland}.

Although Eq. (\ref{wall}) is axisymmetric, as the angular coordinate $\theta
$ does not appear in it, the superposition of the two vortices in Eq. (\ref%
{hybrid}) breaks the isotropy of the pattern in the midplane: $\left\vert
\phi \left( \rho ,\theta ,z=0\right) \right\vert ^{2}=4\phi _{0}^{2}(\rho
)\cos ^{2}\left( S\theta \right) $, where $\phi _{+}\left( \rho ,z=0\right)
=\phi _{-}\left( \rho ,z=0\right) \equiv \phi _{0}(\rho )$. The latter
pattern is close to the numerically found midplane structures, as can be
seen in the right column of Fig. \ref{fig1}.

It is relevant to stress that, unlike the vortical modes of the $S/S$ type
considered above, the vortex-antivortex hybrids cannot be classified as
symmetric or antisymmetric species, with respect to the top and bottom
sections of the ``peanut" profile. Indeed, a rotation of an hybrid state by
angle $\pi /2$ about the vertical axis is effectively tantamount to adding a
phase shift of $\pi $ between the top vortex and the bottom antivortex.

\begin{figure}[t]
%\centering
\includegraphics[width=9cm] {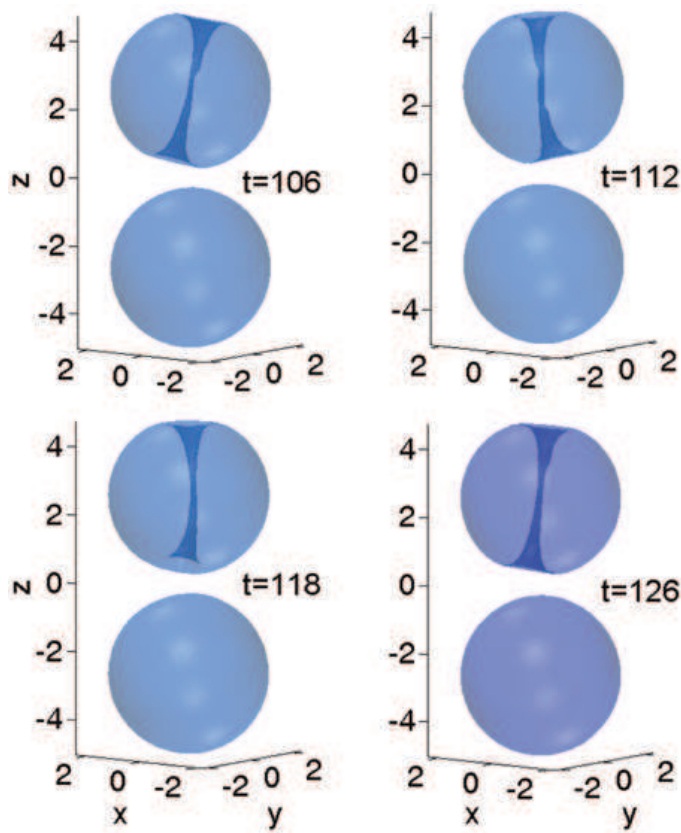}
\caption{(Color online) A generic example of the robust spontaneously
established precession of a top vortex placed above a zero-vorticity base in
the configuration with $d=5$ in Eq. (\protect\ref{sigma}). The vortex and
base components were taken from the respective stable symmetric solutions of
the $S=1/1$ and $S=0/0$ types, with a common value of the chemical
potential, $\protect\mu =15$. The isosurfaces are displayed at density level
$\left\vert \protect\psi \left( x,y,z\right) \right\vert ^{2}=1$. The period
of the steady precession is $\Delta t\approx 20$. }
\label{fig4}
\end{figure}

Another novel type of hybrid modes, which is studied by means of direct
simulations below, is one of the $S=1/0$ type. In this case, the ansatz in
the form of the superposition of the vortical ($S=1$) and fundamental ($S=0$%
) modes in the top and bottom sections of the system [cf. Eq. (\ref{hybrid}%
)] does not lead to a self-consistent approximation. In this situation
simulations reveal robust dynamical regimes, with the vortex \emph{precessing%
} on top of the fundamental soliton, as illustrated in Fig. \ref{fig4}. Our
simulations show that, in suitable parameter regions, such \emph{%
spontaneously established} dynamical states survive over indefinitely long
evolution times (far exceeding $t=100$).

%\begin{widetext}
%\end{widetext}

\section{Numerical results}

\subsection{Stationary modes and their stability}

Stationary solutions for the basic types of 3D confined modes that are
defined above were obtained as solutions of Eq. (\ref{2}), with the
modulation function (\ref{sigma}), by means of the Newton's method. The
stability of the so generated families of different modes was studied by
means of a numerical solution of eigenvalue problem (\ref{eigen}), and
verified by direct simulations of perturbed evolution of the modes that were
performed with the help of the split-step algorithm.

As indicated above, the solution families are naturally represented by
dependences $N(\mu )$, which are collected in Fig. \ref{fig3} for two values
of $d$ in Eq. (\ref{sigma}), \textit{viz}., $d=3$ in (a,b), and $d=5$ in
(c). The plots distinguish stable and unstable families, and include the
analytical results presented above, \textit{viz}., the prediction of the TFA
for the symmetric modes of the $S=0/0$ and $S=1/1$ types (see Eq. (\ref{NTFA}%
)). In addition, the norm difference between the symmetric and antisymmetric
$S=0/0$ states, as predicted analytically by Eq. (\ref{I}), is presented,
together with its numerically computed counterpart, in panel (d) for $d=0$.

Typical examples of all stationary modes are displayed in Fig. \ref{fig1},
their shapes being additionally illustrated by means of vertical cross
sections in Fig. \ref{fig2}. Antisymmetric 3D modes of the $0/0$ and $1/1$
types seem as built of two oblate fundamental solitons or vortices
``levitating" on top of each other. Symmetric $0/0$ and $1/1$ states, which
feature ``peanut"-like shapes, transform into solutions reported in Ref.
\cite{gyro} with the decrease of separation $d$ between the nonlinearity
minima.

A salient finding is the existence of the stationary \textit{hybrid modes},
stable and unstable examples of which are shown, respectively, for $S=1/-1$
and $S=2/-2$. Cross sections of the hybrids are displayed in the right
column of Fig. \ref{fig2}, along the nodal directions in the midplane ($z=0$%
). Such a choice of the presentation is required because, as noted above,
the hybrid modes are axially asymmetric, in contrast to the isotropic ones
of types $0/0$ and $1/1$.

As concerns the stability of the modes, all branches in Fig. \ref{fig3}
satisfy the \textit{anti-Vakhitov-Kolokolov} criterion, $dN/d\mu >0$, which
is a necessary (but, generally, not sufficient) condition for the stability
of self-trapped states supported by repulsive nonlinearities \cite{anti}
(the Vakhitov-Kolokolov criterion per se, $dN/d\mu <0$, is a necessary
condition for the stability of solitons in media with attractive
nonlinearities \cite{VK,Kuznetsov}). While the families of the symmetric
modes of the $S=0/0$ and $S=1/1$ types were found to be completely stable,
only small segments [the bold black ones in Fig. \ref{fig3}(a,b)] of their
antisymmetric counterparts are stable too.

\begin{figure}[t]
\centering\centering\includegraphics[width=8cm]{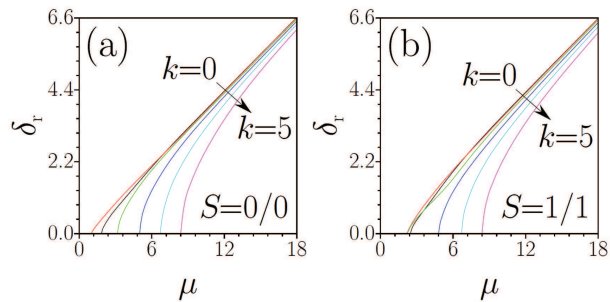}
\caption{(Color online) Instability growth rates for antisymmetric modes
with $S=0/0$ (a) and $S=1/1$ (b), versus the chemical potential of the
unperturbed state, and integer azimuthal index ($0\leq k\leq 5$) of the
perturbation eigenmode, defined as per Eq. (\protect\ref{pert}). The most
destructive perturbations at small values of $\protect\mu $ correspond to $%
k=1$ [the red curves in panels (a) and (b)]. Stable are regions at small
values of $\protect\mu $, where $\protect\delta _{\mathrm{r}}=0$.}
\label{fig5}
\end{figure}

The stability-instability transition for the antisymmetric $0/0$ and $1/1$
states at small values of $\mu $ is additionally illustrated by Fig. \ref%
{fig5}, which displays the instability growth rates, $\delta _{\mathrm{r}%
}\equiv \mathrm{Re}(\delta )$, as functions of $\mu $ and azimuthal index $k$
(limited to $k\leq 5$), see Eqs. (\ref{pert}) and (\ref{eigen}). In
particular, an unusual peculiarity is that, for the antisymmetric state of
the $0/0$ type, the dominant instability mode for small $\mu $ corresponds
to $k=1$ [the red curve in Fig. \ref{fig5}(a)], while zero-vorticity states
are normally destabilized solely by perturbations with $k=0$ \cite{review1}.
These instability eigenvalues are complex, hence the respective dynamics is
oscillatory (see below).

Another important finding is a large stability region of the hybrid modes
with $S=1/-1$, as shown, in Fig. \ref{fig3}(c), for $d=5$ in Eq. (\ref{sigma}%
). It is worthy to note that this stability region strongly depends on $d$:
a detailed analysis reveals that the vortex-antivortex hybrids are
completely unstable at $d\leq 4$, when the vortex and antivortex
constituents of the hybrid are relatively strongly pressed onto each other,
and a stability region appears at $d>4$, being $\mu \leq 15.8$, i.e., $%
N<394.9$, at $d=4.5$, and $\mu \leq 13.5$, i.e., $N<329.3$, at $d=5$. Thus,
it is worthy to note that the size of the stability region does not grow
monotonously with the increase of $d$.

\subsection{Dynamical states: The evolution of unstable modes, and robust
precessing hybrids with $S=1/0$}

\begin{figure}[t]
\centering\centering\includegraphics[width=8cm]{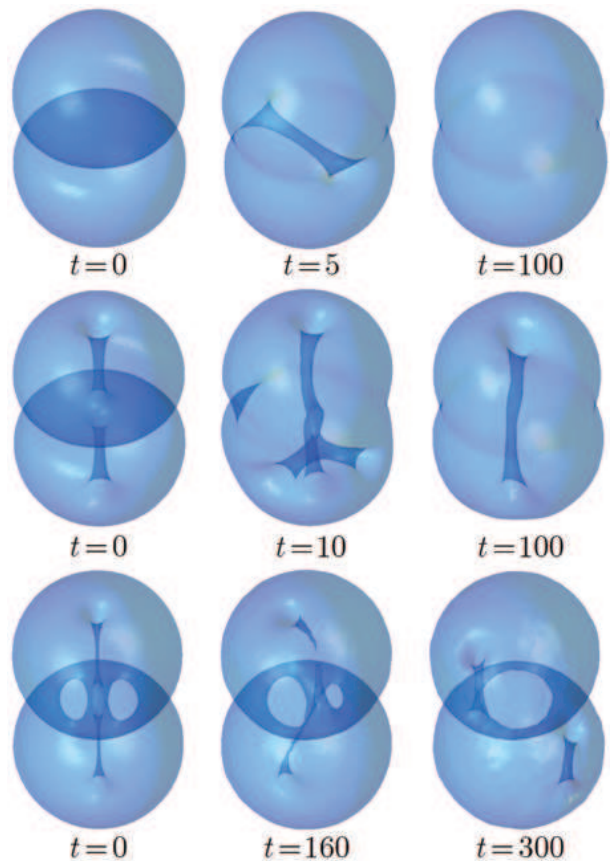}
\caption{(Color online) Generic examples of the evolution of unstable
antisymmetric modes with $S=0/0$ and $S=1/1$ [the top and middle rows,
respectively; they correspond to circles on red curves in Figs. \protect\ref%
{fig7}(a) and (b), with $\protect\mu =5$ and $d=3$]: spontaneous
transformation into the respective symmetric modes. Note that the vortical
structure\emph{\ survives}, in the case of $S=1/1$, although the instability
is strong. The bottom row: the spontaneous transformation of an unstable
hybrid with $S=1/-1$ [at $\protect\mu =17$, $d=5$, which corresponds to the
circle on the red segment in Fig. \protect\ref{fig3}(c)] into a fundamental
symmetric mode, with zero vorticity. The isosurfaces are displayed at
density level $\left\vert \protect\psi \left( x,y,z\right) \right\vert
^{2}=0.1$ in the top and middle rows, and at the level of $\left\vert
\protect\psi \left( x,y,z\right) \right\vert ^{2}=0.5$ in the bottom row.}
\label{fig6}
\end{figure}

Typical examples of the evolution of perturbed modes, of those types which
may be unstable [they are marked by circles on red branches in Fig. \ref%
{fig3}(a,b,c)], are displayed in Fig. \ref{fig6}. In all the cases, the
evolution keeps initial values of the norm and angular momentum (\ref{M}).
In particular, weakly unstable antisymmetric (dipole) modes with $S=0/0$ and
$S=1/1$, which are taken close to the boundary of the stability region [see
Figs. \ref{fig3} (a,b) and \ref{fig5}], feature only small oscillations of
their amplitude, while keeping their dipole structure and vorticity (in the
case of $S=1/1$). That is, the regions of \textit{effective stability} for
the dipole modes are actually larger than the rigorously defined bold black
segments on the respective $N(\mu )$ curves in Figs. \ref{fig3}(a,b). On the
other hand, at greater values of $N$, stronger instability destroys the
dipole structure, tending to transform the antisymmetric modes into their
symmetric counterparts, as shown in the top and middle rows of Fig. \ref%
{fig6}.

A remarkable feature of the instability-induced evolution (well corroborated
by the simulations) is that the vortical structure survives in the course of
the spontaneous transformation of the unstable dipole mode of the $1/1$ type
into its stable symmetric counterpart (see the middle row in Fig. \ref{fig6}%
). As concerns unstable hybrids, they, quite naturally, exhibit spontaneous
annihilation of the vortex with antivortex, thus gradually transforming
themselves into symmetric zero-vorticity (fundamental) states, as seen in
the bottom row of Fig. \ref{fig6}. On the other hand, stable hybrid solitons
do not show any conspicuous shape transformations even at $t>10^{3}$, and
even in the presence of strong initial perturbations.

As indicated above, hybrids with $S=1/0$, built of a vortex placed on top of
a fundamental mode, cannot form a stationary state. Nevertheless, direct
simulations show, as shown in Fig. \ref{fig4}, that the hybrids of this type
readily self-trap in a dynamical form, with the vortex performing periodic
precession above the zero-vorticity base. The respective initial
configuration was constructed by juxtaposing the top and bottom components
taken as respective parts of the symmetric vortex and fundamental states,
with $S=0/0$ and $S=1/1$, which were preliminarily generated, for equal
values of the chemical potential, in the same trapping configuration. A
systematic numerical analysis shows that such robust dynamical regime is
observed in a broad parametric area, provided that $d$ is not too small,
namely, $d\geq d_{\min }\approx 4.8$.

\section{Conclusions}

Using a systematic numerical analysis and a range of analytical
approximations, we have discovered several previously unknown species of
self-trapped complex 3D field states, that are supported by the local
strength of a repulsive cubic nonlinearity growing from two local minima to
the periphery, along the axial and radial directions alike. We have shown
that the corresponding axisymmetric ``peanut"-shaped 3D
nonlinearity-modulation profiles support families of vortex states, which
are both symmetric and antisymmetric with respect to the top-bottom
reflection. The same system gives rise to a novel species of stable
stationary top-bottom vortex-antivortex hybrids, which was not reported
previously in any 3D setting, to the best of our knowledge. Another newly
found species of self-trapped robust dynamical hybrid states exhibits stable
precession of a top vortex above a bottom fundamental mode. In addition, we
showed (in the Appendix) that systems with ``bead"-shaped 3D modulation
profiles produce the first example of exact analytical solutions for stable
3D vortex solitons. Settings of such type may be realized in media that
allow a local control of the cubic self-repulsive nonlinearity by means of
external fields. In particular, this is possible in Bose-Einstein
condensates, using the Feshbach resonance controlled by appropriately
designed nonuniform magnetic or optical fields. The latter settings suggest
a physical realization of the predicted self-trapped modes.

\begin{acknowledgments}
B.A.M. appreciates hospitality of ICFO. The work of R.D. and B.A.M.
was supported, in a part, by the Binational (US-Israel) Science
Foundation through grant No. 2010239, and by the German-Israel
Foundation through grant No. I-1024-2.7/2009. R.D. and T.M.
acknowledge support provided by the Deutsche Forschungsgemeinschaft
(DFG) via the Research Training Group (GRK) 1464, and computing time provided by
PC$^2$ (Paderborn Center for Parallel Computing). YVK and LT are
supported by the Severo Ochoa Excellence program of the Government
of Spain.
\end{acknowledgments}

\section*{Appendix: Exact solutions for 3D vortex modes}

None of the models studied above in this work or elsewhere have produced an
exact analytical solution for 3D vortex solitons (there is a method which
makes it possible to construct exact solutions of NLS/GP equations with
variable coefficients which are deliberately designed as an explicit
coordinate transformation of the 1D integrable equation \cite{similarity},
but we here aim to produce truly three-dimensional solutions). Here, as a
direct proof of the existence of such modes, we address an additional model,
with a ``bead"-shaped modulation structure, which produces exact solutions
for 3D vortices. It is based on the following equation, written, as Eq. (\ref%
{sigma}), in the cylindrical coordinates:
\begin{equation}
i\frac{\partial \psi }{\partial t}=-\nabla ^{2}\psi +\left( 1+\frac{z^{2}}{%
b^{2}\rho ^{2}}\right) \exp \left( \frac{1}{2}\left( z^{2}+b\rho ^{2}\right)
\right) \left\vert \psi \right\vert ^{2}\psi .  \label{GPE}
\end{equation}%
While constant $b>0$ controls the anisotropy of the modulation profile, the
singularity of the self-repulsion strength in the pre-exponential factor at $%
\rho =0$ may be created in BEC by means of a control field which attains the
exact Feshbach resonance on the axis (at $\rho \rightarrow 0$), as well as
at $\rho \rightarrow \infty $.

\begin{figure}[t]
\centering\centering\includegraphics[width=8cm]{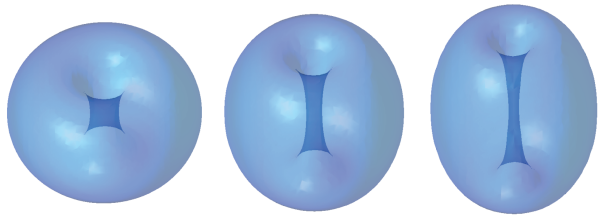}
\caption{(Color online) Isosurface plots at density level $\left\vert
\protect\psi \left( x,y,z\right) \right\vert ^{2}=0.2$ display the shape of
\emph{exact solutions} given by Eq. (\protect\ref{U}) for stable confined
vortices in the model based on Eq. (\protect\ref{GPE}) with the bead-shaped
modulation structure. The anisotropy parameter is $b=1$ (left), $2$
(center), $3$ (right).}
\label{fig7}
\end{figure}

An exact 3D solution to Eq. (\ref{GPE}), which produces a confined vortex
with topological charge $1$, is
\begin{equation}
\psi =\frac{b}{2}\rho e^{i\theta -i\mu t}\exp \left( -\frac{1}{4}\left(
z^{2}+b\rho ^{2}\right) \right) ,  \label{U}
\end{equation}%
with chemical potential $\mu =\left( 1+4b\right) /2$ and norm $N=\sqrt{2}\pi
^{3/2}$ (note that the norm does not depend on $b$). This is a particular
solution belonging to a family of vortex solitons, which, in the general
form, can be constructed by means of numerical methods (not shown here).
Examples of the exact vortices, for different values of the anisotropy
parameter $b$, which are displayed in Fig. \ref{fig7}, indeed feature
bead-like shapes. The computation of the stability eigenvalues and direct
simulations demonstrate that the exact vortex solutions are \emph{stable}.

\end{document}